\documentclass[twocolumn,showpacs,preprintnumbers,amsmath,amssymb,prl]{revtex4}

\usepackage{graphicx}
\usepackage{bm}

\begin{document}

\title{Direct evidence for strong crossover of collective excitations and positive sound dispersion in the supercritical state}

\author{Yu. D. Fomin, V. N. Ryzhov, E. N. Tsiok, V. V. Brazhkin}
\affiliation{Institute for High Pressure Physics, Russian Academy of Sciences, Troitsk 142190, Moscow, Russia}
\author{K. Trachenko}
\affiliation{School of Physics and Astronomy, Queen Mary University of London, Mile End Road, London E1 4NS, United Kingdom}

\begin{abstract}
Supercritical state has been viewed as an intermediate state
between gases and liquids with largely unknown physical
properties. Here, we address the important ability of
supercritical fluids to sustain collective excitations. We
directly study propagating modes on the basis of correlation
functions calculated from extensive molecular dynamics
simulations, and find that the supercritical system sustains
propagating solid-like transverse modes below the Frenkel line but
becomes devoid of transverse modes above the line where it
supports longitudinal modes only. Important thermodynamic
implications of this finding are discussed. We directly detect
positive sound dispersion (PSD) below the Frenkel line where
transverse modes are operative, quantitatively explain its
magnitude on the basis of transverse and longitudinal velocities.
PSD disappears above the Frenkel line which therefore demarcates
the supercritical phase diagram into two areas where PSD does and
does not operate.
\end{abstract}

\maketitle

Most important thermodynamic and dynamical properties of an
interacting system are related to collective excitations. This has
been widely ascertained in solids and, to a smaller degree, in
liquids \cite{march1,balu,hansenmcd}. In contrast, there is nearly
no discussion and understanding of what kind of collective modes
propagate in the supercritical state of matter, how they evolve
with temperature and what are the physical consequences. This is
related to the lack of understanding of the supercritical state
itself \cite{sup1,sup2}. In existing view, no difference can be
made between gases and liquids above the critical point. At the
same time, it was long suspected that the supercritical state can
show elements of gas and liquid behavior depending on pressure and
temperature. We have recently addressed this point
\cite{frprl,frpre,ufn} and proposed that a distinction between
gas-like and liquid-like states can be formulated in a rigorous
sense. We have put forward the concept of the Frenkel line (FL)
demarcating the gas-like and liquid-like behavior at arbitrarily
high pressure and temperature above the critical point. Crossing
the line results in the qualitative change of particle dynamics:
particle motion is purely diffusive above the line as in a gas,
and combines diffusive and oscillatory motion below the line as in
a liquid. Mathematically, the FL can be defined by the
disappearance of the minima of the velocity auto-correlation
function on temperature increase (or pressure decrease),
corresponding to the gas-like monotonic decay \cite{frprl}.

At first glance, the above qualitative change of supercritical
system behavior is dynamical only. Yet crossing the Frenkel line
should also have an important thermodynamic consequence, through
the pronounced crossover of collective excitations. Indeed, the
crossover from oscillatory-diffusive to purely diffusive dynamics
corresponds to $\tau=\tau_{\rm D}$ \cite{frpre}, where $\tau$ is
liquid relaxation time and $\tau_{\rm D}$ is the minimal period of
vibration of transverse waves. Combining this with the proposal
that a liquid sustains propagating transverse waves with frequency
above $\frac{1}{\tau}$, the FL corresponds to the complete loss of
two transverse modes at all available frequencies. The crossover
at the FL corresponds to specific heat $c_v=2k_{\rm B}$, which
includes the contribution of the energy of one longitudinal mode,
$NT$, and kinetic terms of two transverse modes of $\frac{Nk_{\rm
B}T}{2}$ each \cite{phonon1}, where $N$ is the number of
particles. The state where $c_v=2k_{\rm B}$ is expected to be
accompanied by the crossover of $c_v$ to the state with only one
longitudinal mode remaining, the prediction supported by the
numerical results \cite{phonon2}. Consistent with the above
discussion, the disappearance of the minima of the velocity
auto-correlation function (VAF) coincides with $c_v=2k_{\rm B}$
\cite{frprl}. This strongly suggests that the dynamical crossover
at the FL is intimately related to an important change of
thermodynamical properties of the supercritical system related to
the crossover of collective excitations.

In view of the current interest in the supercritical state and
importance of collective excitations, it is essential to follow
their evolution directly across the FL and discuss physical
implications. This is the primary goal of this paper. For the
first time, we directly study collective excitations in the
supercritical state on the basis of correlation functions
calculated from molecular dynamics simulations and discuss
important implications for key thermodynamic properties such as
energy and heat capacity. We then turn to interesting dynamical
phenomena such as PSD. We directly detect PSD below the FL where
transverse modes operate but not above the FL, and provide a
quantitative description of the effect.

\begin{figure}
\includegraphics[width=7cm, height=7cm]{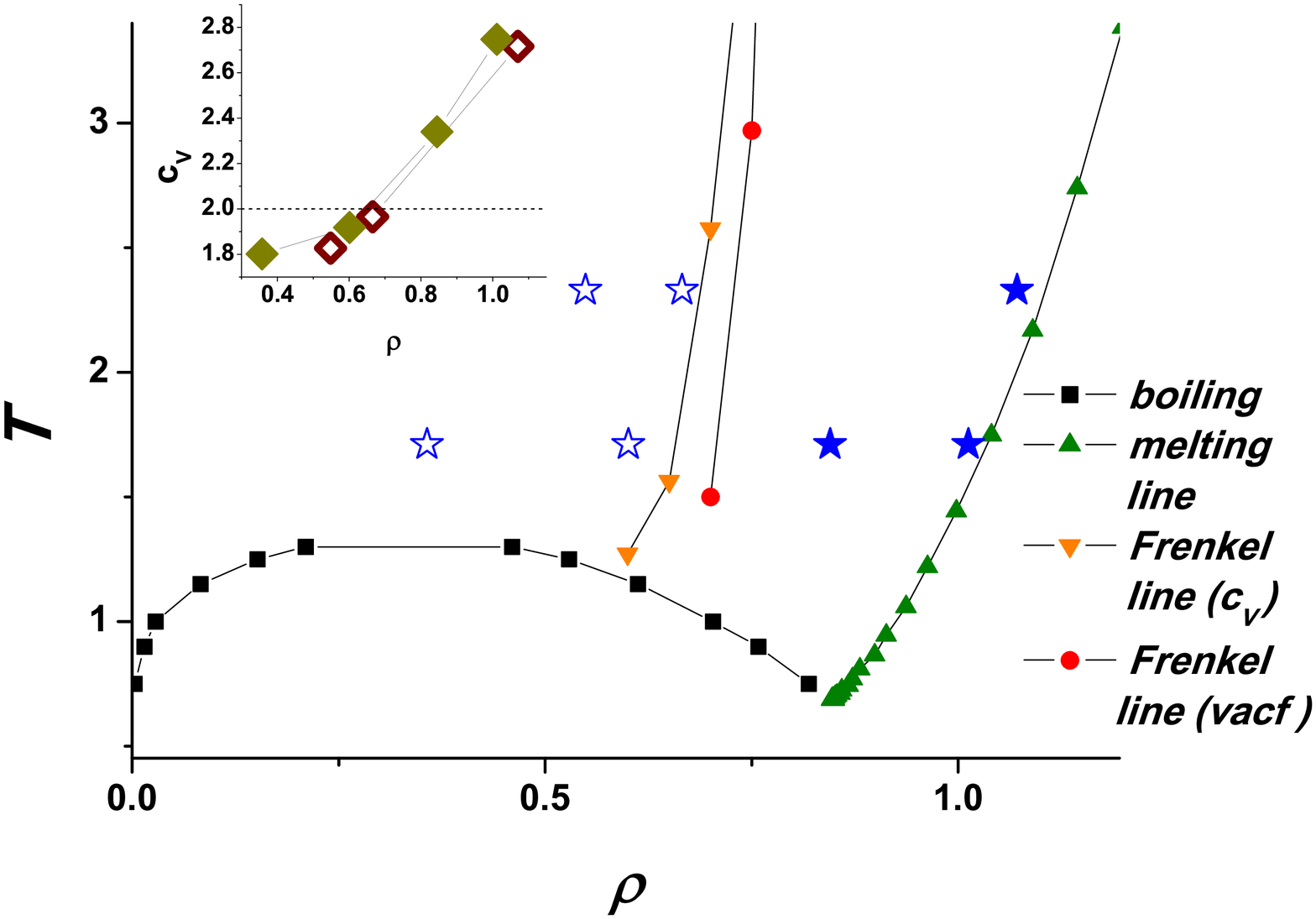}
\caption{The location of points studied in the present work
(stars) on the phase diagram. Full and open symbols correspond to
the rigid liquid below the FL and non-rigid gas-like fluid above
the FL. The Frenkel line obtained from VAFs and from $c_V=2k_B$
rules are also shown. The inset shows specific heat $c_v$ along
the two isotherms studied. Filled and open diamonds correspond to
$T=1.71$ and $T=2.33$, respectively.} \label{points}
\end{figure}

We have performed MD simulations of the commonly used
Lennard-Jones (LJ) system with parameters $\varepsilon=119.8$ K
and $\sigma=3.405$ \AA\ corresponding to Argon. We have used
$4000$ particles in the cubic box with periodic boundaries. The
system was equilibrated for $1.0 \cdot 10^6$ time steps, followed
by $0.5 \cdot 10^6$ production steps. Nose-Hoover and
constant-energy ensembles were used for equilibration and
production runs, respectively. The time step was $0.001$ in LJ
units. We have simulated several temperature and pressure points
in the supercritical state below and above the FL. Accordingly, we
have simulated three densities at $T=2.33$: $\rho=0.549$, $0.6655$
and $1.071$ (here and below, $T$ and $\rho$ are in LJ units) and
four densities at $T=1.71$: $\rho=0.357$, $0.601$, $0.845$ and
$1.012$. The point ($\rho=0.6655$, $T=2.33$) is very close to the
Frenkel line while the points ($\rho=0.549$, $T=2.33$) and
($\rho=1.071$, $T=2.33$) are above and below the FL, respectively.
The location of points under investigation is shown in Fig.
~\ref{points}. We refer to two regimes of liquid dynamics above
and below the FL as ``non-rigid'' gas-like fluid and ``rigid''
liquid, reflecting the system ability to sustain solid-like
transverse modes.

\begin{figure}
\includegraphics[width=7cm, height=6cm]{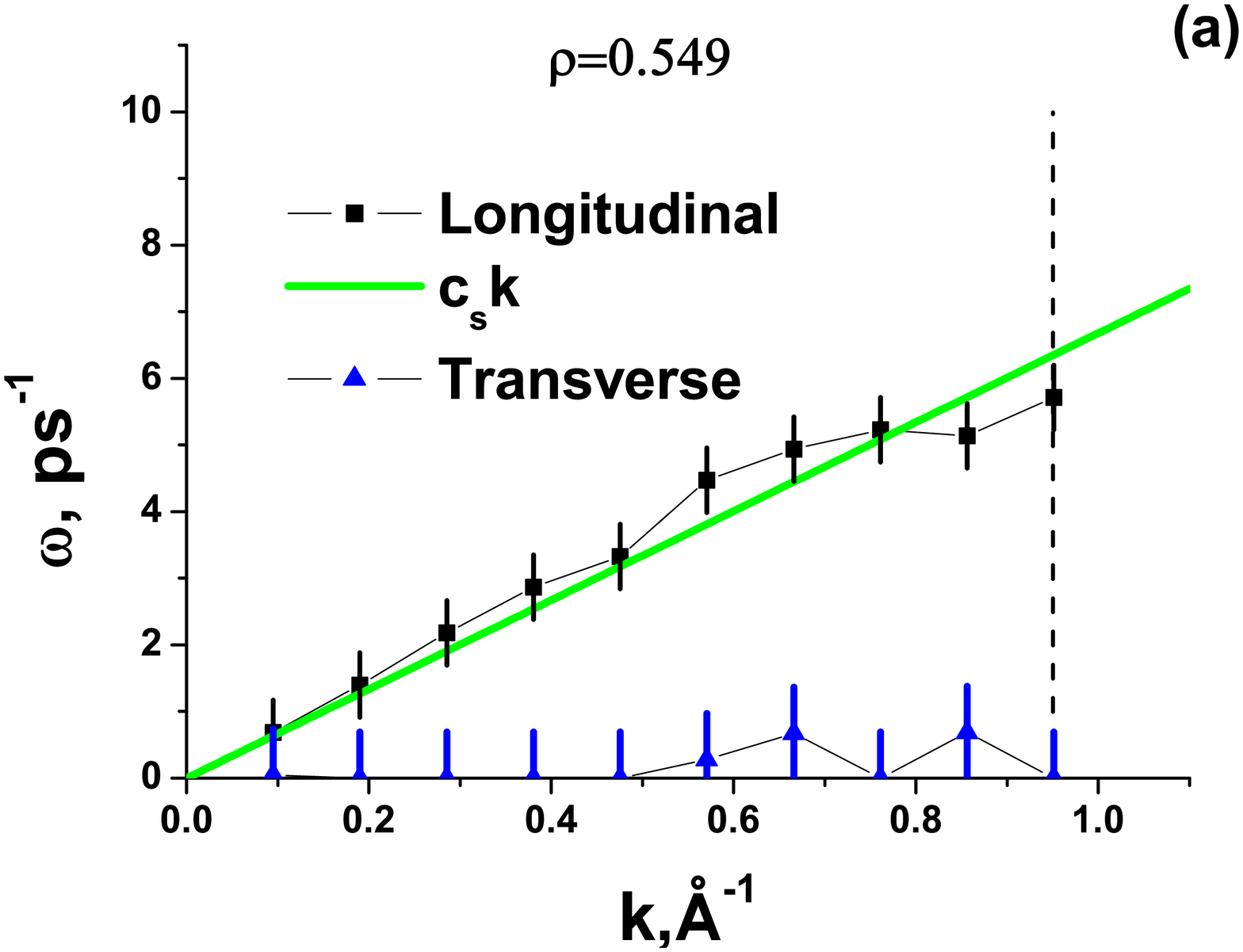}
\includegraphics[width=7cm, height=6cm]{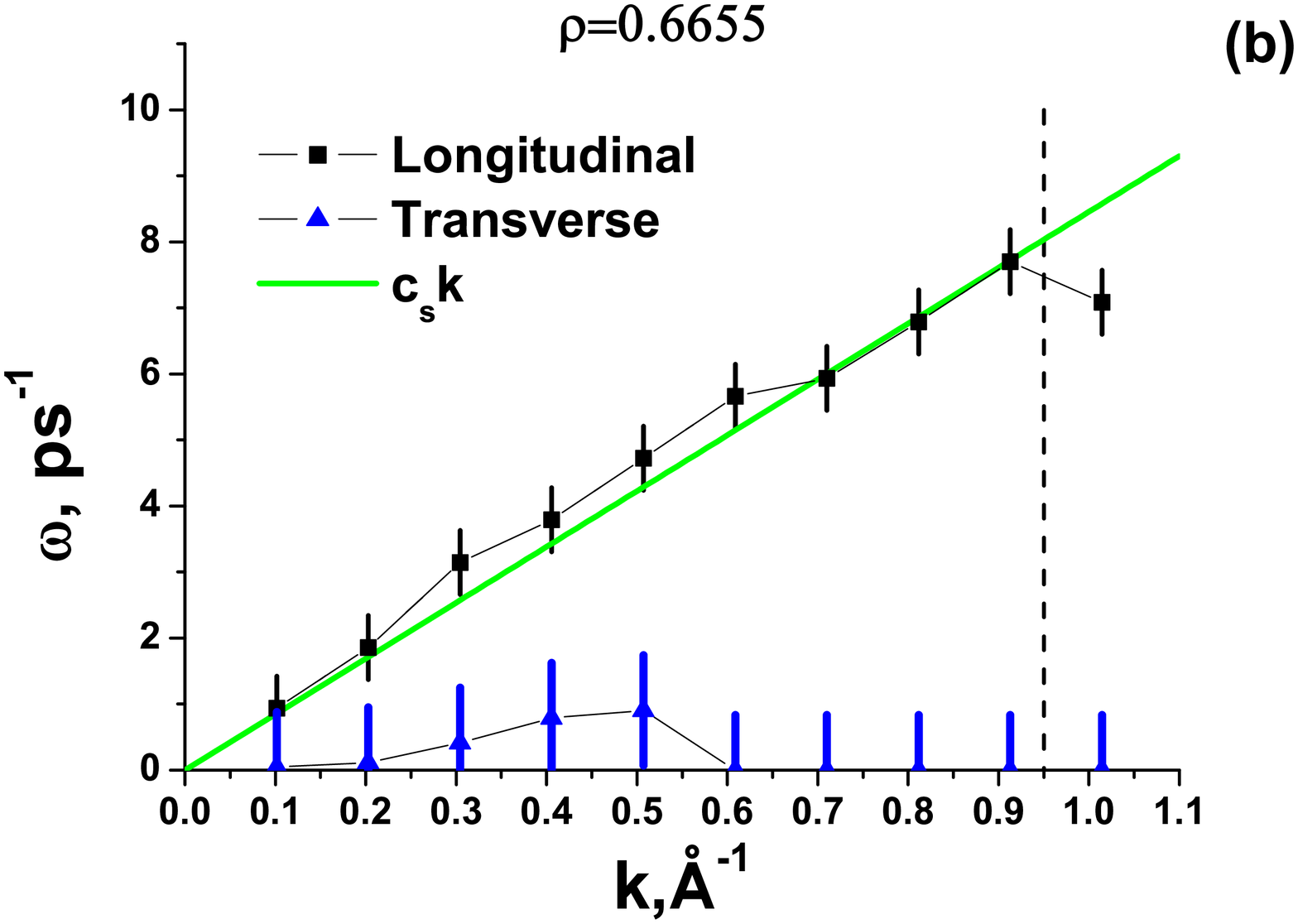}
\includegraphics[width=7cm, height=6cm]{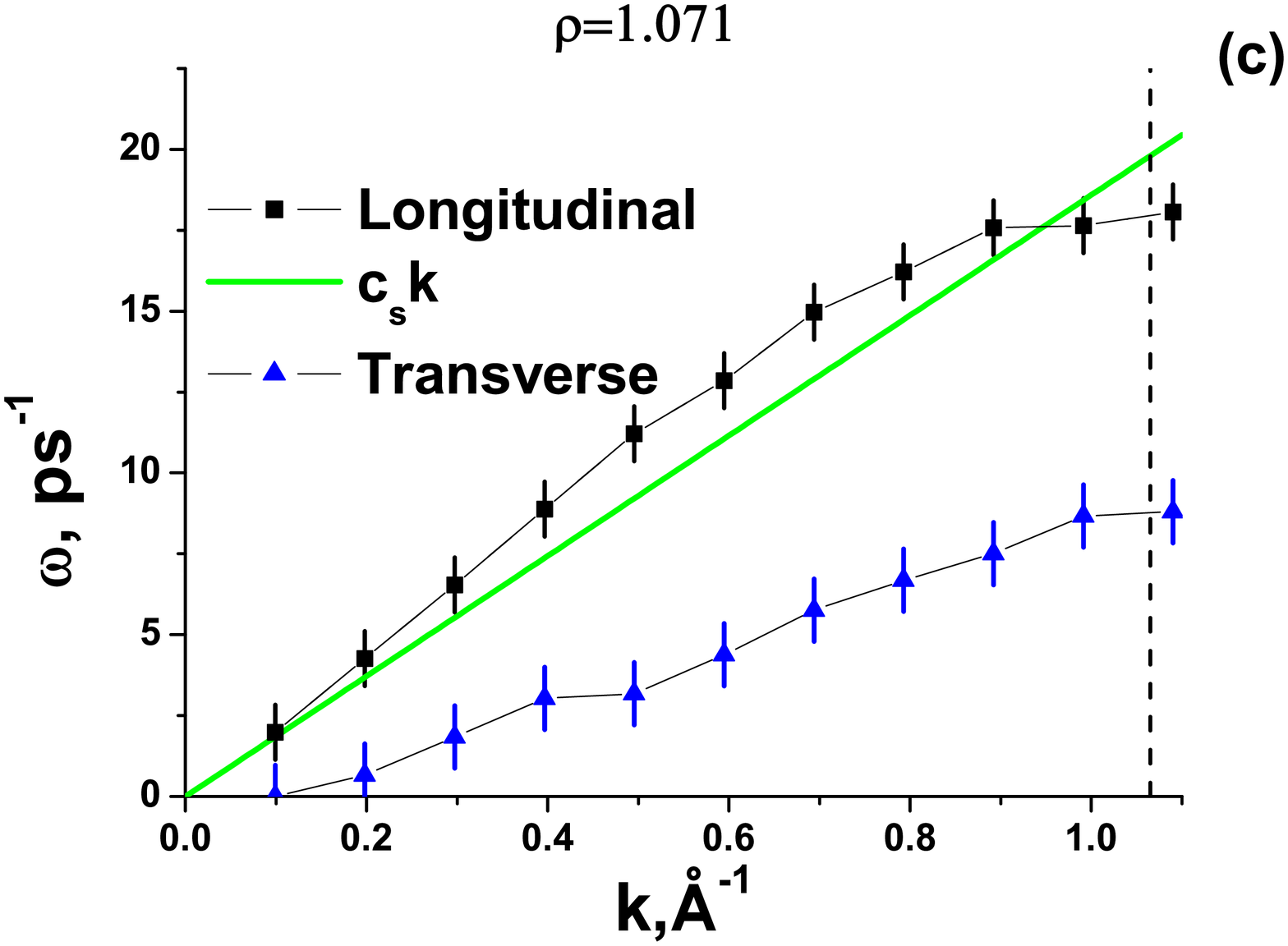}
\caption{Dispersion curves for $T=2.33$ at three densities
studied. The dashed vertical lines mark the boundary of
pseudo-Brillouin zone. The speed of sound $c_s$ for Ar is from the
NIST database.} \label{sp-t233}
\end{figure}

To directly address the evolution of collective modes, we have
studied the dispersion of longitudinal and transverse modes and
have calculated the correlation functions
$C_L(k,t)=\frac{k^2}{N}\langle J_z({\bf k},t) \cdot J_z(-{\bf
k},t)\rangle$ and $C_T(k,t)=\frac{k^2}{2N} \langle J_x({\bf
k},t)\cdot J_x(-{\bf k},t)+J_y({\bf k},t) \cdot J_y(-{\bf
k},t)\rangle$, where $J({\bf k},t)=\sum_{j=1}^N {\bf v}_j
e^{-i{\bf k r}_j(t)}$ is the velocity current \cite{hansenmcd}.
Dispersion curves of longitudinal and transverse excitations are
obtained from the location of maxima of Fourier transforms
$\tilde{C}_L({\bf k},\omega)$ and $\tilde{C}_T({\bf k},\omega)$
respectively. The spectra for $T=2.33$ at three densities studied
are shown in Figs. ~\ref{sp-t233} (a)-(c), together with the
dispersion curves of both longitudinal and transverse excitations.

We observe that in the non-rigid gas-like fluid at $\rho=0.549$,
the dispersion curve is close to the Debye approximation $\omega =
c_s k$, where $c_s$ is the adiabatic speed of sound. Small upward
deviation from the linear dependence (positive sound dispersion,
PSD) is observed, but is within the simulation error bars (the
errors that include the calculation of the correlation functions
$C_L(k,t)$ and $C_T(k,t)$ and taking their Fourier transforms). We
also observe that the transverse excitations are absent at this
density within the error. The same behavior is seen at
$\rho=0.6655$ very close to the FL. A notable difference in
behavior appears at $\rho=1.071$ which corresponds to the rigid
liquid below the FL: we clearly observe PSD in case of
longitudinal modes and a distinct transverse branch for the
transverse ones.

\begin{figure}
\includegraphics[width=7cm, height=6cm]{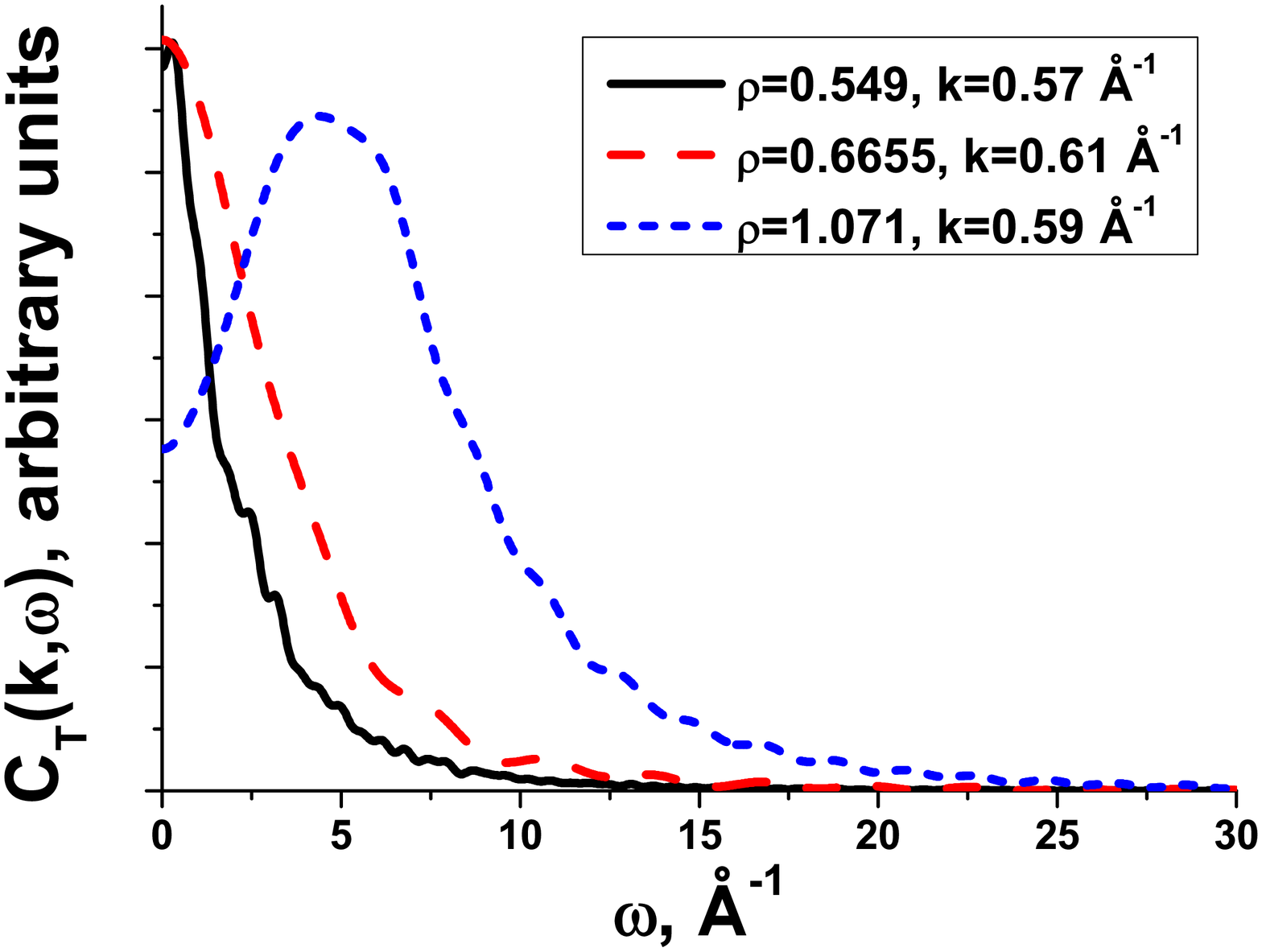}
\caption{$\tilde{C}_T(k,\omega)$ at $T=2.33$ at three densities
studied, showing the peak and propagating transverse excitations
at high density and their absence at low density.} \label{ct-3den}
\end{figure}

In order to emphasize the qualitative change of the spectrum of
transverse excitations on crossing the FL, we show examples of
functions $\tilde{C}_T(k,\omega)$ for all three densities at
approximately the same value of $k$ (Fig. ~\ref{ct-3den}). We
observe that no peak is found in the current correlation function
at $\rho=0.549$ and $0.6655$, while the clear peak is seen at
$\rho=1.071$ signifying the propagating transverse mode with
frequency $\omega=4.38$ ps$^{-1}$.

\begin{figure}
\includegraphics[width=7cm, height=6cm]{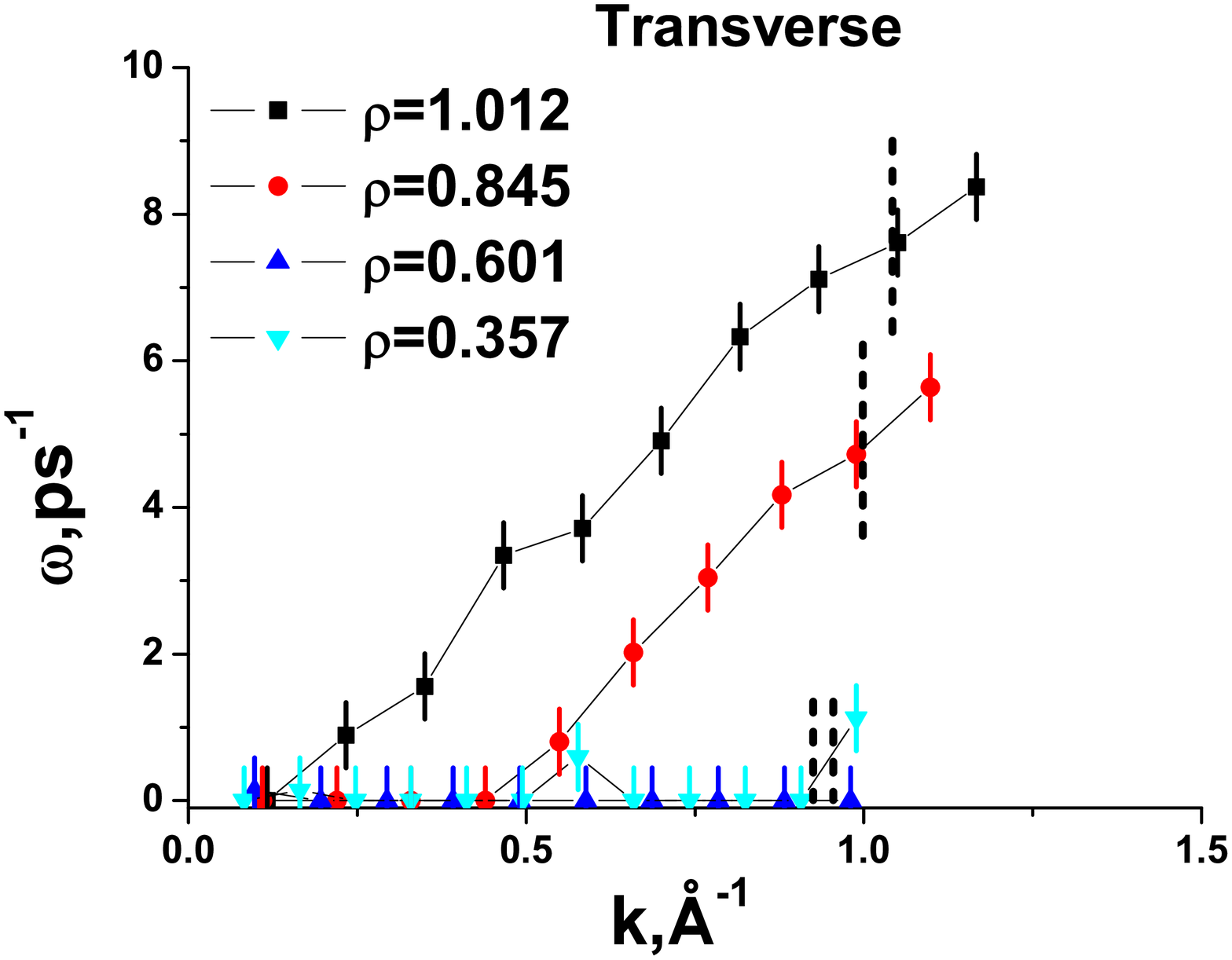}
\caption{Spectra of transverse excitations at $T=1.71$ at four
densities. Short dashed lines mark the boundary of the first
pseudo-Brillouin zone.} \label{trans-t171}
\end{figure}

The same picture of evolution of transverse modes emerges from the
simulation at the different temperature, $T=1.71$. According to
Fig. \ref{trans-t171}, no transverse excitations are found above
the FL at $\rho=0.357$ and $0.601$. At higher densities below the
FL, the transverse excitations are clearly seen: at $\rho=0.845$
and $\rho=1.012$, they start from $k>0.44$ and from the origin,
respectively.

Our results of direct calculation of transverse modes have two
important implications for understanding the thermodynamic
behavior of the supercritical state. First, the theoretical
calculation implies that the disappearance of transverse waves
gives $c_v=2k_{\rm B}$ \cite{phonon1} as discussed above. Combined
with the directly calculated evolution of transverse waves and
their disappearance at the FL, this implies that $c_v$ of the
supercritical system is $2k_{\rm B}$ at the FL. This value is
consistent with the $c_v$ calculated from our MD simulations at
the FL, as was the case in the previous work \cite{frprl}. Second,
the disappearance of transverse modes at the FL implies that the
evolution of the supercritical system above the FL is governed by
the remaining longitudinal mode only, with the particle mean free
path defining the minimal wavelength possible. This is a mechanism
different from that governing the system vibration energy below
the FL. Therefore, we expect to find the crossover of $c_v$ at the
FL between two different behaviors. Such a crossover was seen in
the previous MD simulations \cite{phonon2}.


In addition to thermodynamic implications above, the disappearance
of transverse waves at the FL has an important consequence for the
phenomenon of PSD mentioned above. The physical origin of the PSD
has remained controversial, including understanding relative
contributions of the above Frenkel mechanism and other effects
such as disorder. Experimentally, PSD has been well-studied in
viscous liquids where the system starts sustaining rigidity at MHz
frequencies (see, e.g., Ref. \cite{jeong}). Here, PSD is seen at
fairly large wavelengths at which the liquid can be considered as
a homogeneous medium. It is generally agreed \cite{jeong} that in
this range of frequencies, PSD originates from the Frenkel
mechanism: at high frequency where liquid shear response and shear
modulus become non-zero, the propagation velocity crosses over
from its hydrodynamic value $v=\sqrt{\frac{B}{\rho}}$ to the
solid-like elastic value $v=\sqrt{\frac{B+\frac{4}{3}G}{\rho}}$,
where $B$ and $G$ are bulk and shear moduli, respectively. At
smaller wavelengths approaching the length of medium and
short-range order, the wave feels structural inhomogeneities, and
disorder of liquids and glasses starts to affect the dispersion
relationship. PSD, with the relative magnitude of few percent, was
observed in a model harmonic glass and attributed to the
``instantaneous relaxation'' due to fast decay and dissipation of
short-wavelength phonons in a disordered system \cite{fast1}.
Later work demonstrated that starting from mesoscopic wavelengths,
the effective speed of the longitudinal sound can both increase
and decrease \cite{fast2,dec1,dec2}. Different mechanisms and
contributions to PSD were subsequently discussed
\cite{ruocco,fast3}. The instantaneous relaxation is likely to be
significant close to the zone boundary \cite{ruzi}, although large
PSD in silica glass may be related to the effect of mixing with
the low-lying optic modes. In water, PSD was discussed on the
basis of coupling between the longitudinal and transverse
excitations, and it was found that the onset of transverse
excitations coincides with the inverse of liquid relaxation time
\cite{water-fast,cunsolo}, as predicted by Frenkel.

Our finding of the presence and absence of transverse modes below
and above the FL has two important implications for understanding
the phenomenon of PSD. First, we see in Fig. \ref{sp-t233} that
significant PSD does not exist at small density and where no
transverse modes propagate, but emerges in the state where
transverse modes kick in at higher density. This is exactly the
behavior predicted by the Frenkel mechanism above.


Second and perhaps more importantly, we can quantitatively address
the magnitude of the PSD in the supercritical state. A well-known
result for solids is that longitudinal and transverse velocities
can be computed as $v_l=\sqrt{\frac{B+\frac{4}{3}G}{\rho}}$ and
$v_t=\sqrt{\frac{G}{\rho}}$ correspondingly where $G$ is the shear
modulus and $B$ is the bulk one. The same expressions can
be applied to liquids which sustain shear stress at high frequency in the solid-like regime \cite{dyre}.
Combining the expressions for $v_h=\sqrt{\frac{B}{\rho}}$, $v_t$ and $v_l$, we write

\begin{equation}
v_l^2=v_h^2+\frac{4}{3}v_t^2
\label{speeds}
\end{equation}

Using our MD data, we have taken $v_l$ and $v_t$ from the
dispersion curves at $k$ points where the observed PSD is maximal
and where $\omega(k)$ is in the quasi-linear regime before
starting to curve at large $k$. We have used two points on the
phase diagram: $(\rho,T)=(1.071,2.33)$ and $(1.012,1.71)$. The
velocities were taken at $k=0.595$ \AA$^{-1}$ for the former and
$k=0.47$ \AA$^{-1}$ for the latter point. Using the observed
$v_h$, we have calculated $v_l$ using Eq. (\ref{speeds}).
Comparing the actual observed $v_l$ with the calculated one in
Table \ref{table}, we find that the Frenkel mechanism describes
the PSD well.

\begin{table}[ht]
\begin{tabular}{l l l l l l}
\hline
       &    & $v_h$  & $v_t$  & $v_l$ (obs.) & $v_l$ (calc.) \\
       &    & [m/s]  & [m/s]  &[m/s]          &[m/s] \\
\hline
&LJ1  & 1860  & 750$\pm 90$   & 2110$\pm 60$  & 2060$\pm 70$  \\
&LJ2  & 1590  &735$\pm 90$    & 1870$\pm 60$  & 1805$\pm 70$ \\
&Fe   & 3800  & 1870$\pm 50$  & 4370$\pm 30$ & 4370$\pm 50$  \\
&Cu   & 3460  & 1510$\pm 50$  & 3890$\pm 30$  & 3875$\pm 50$  \\
&Zn   & 2780  & 1620$\pm 50$  & 3330$\pm 30$  & 3350$\pm 50$  \\
&Sn   & 2440  & 1220$\pm 150$ & 2890$\pm 30$  & 2820$\pm 150$ \\
\hline
\end{tabular}
\caption{Comparison of predicted and observed $v_l$. The data for
LJ systems is from our current MD simulations. LJ1 corresponds to
($\rho =1.071$, $T=2.33$) and LJ2 to ($\rho=1.012$, $T=1.71$). The
data for Fe, Cu, Zn and Sn are from Refs. \cite{sn,hoso3}.}
\label{table}
\end{table}

We have complemented Table \ref{table} by the data for $v_l$,
$v_t$ and $v_h$ from recent experiments for Fe, Cu, Zn and Sn
\cite{sn,hoso3}. As before, we have used $k$ points where the
observed PSD is maximal and where $\omega(k)$ is in the
quasi-linear regime. For Fe, Cu, Zn, we consider the following $k$
points: $k=7.7$ nm$^{-1}$, (first point on the transverse branch),
$k=7.8$ nm$^{-1}$ (second point on the branch) and $k=8$ nm$^{-1}$
(second point on the branch). For Sn, large PSD is seen at about
$k=3.3$ nm$^{-1}$ at the second point on the longitudinal branch.
To find $v_t$ at this $k$, we extrapolated the higher-lying
transverse points to lower $k$ while keeping them parallel to the
simulation points, yielding $v_t=1220$ m/s. Similar to LJ systems,
we observe perfect quantitative agreement between calculated and
observed $v_l$ in Table \ref{table}.

Coupled with the direct evidence of transverse modes below the FL,
the quantitative agreement between predicted and observed $v_l$
implies that the Frenkel mechanism of PSD quantitatively accounts
for both simulated and experimental data in the wide range of $k$
spanning at least a half of the first pseudo-Brillouin zone.

\begin{figure}
\includegraphics[width=5cm, height=5cm]{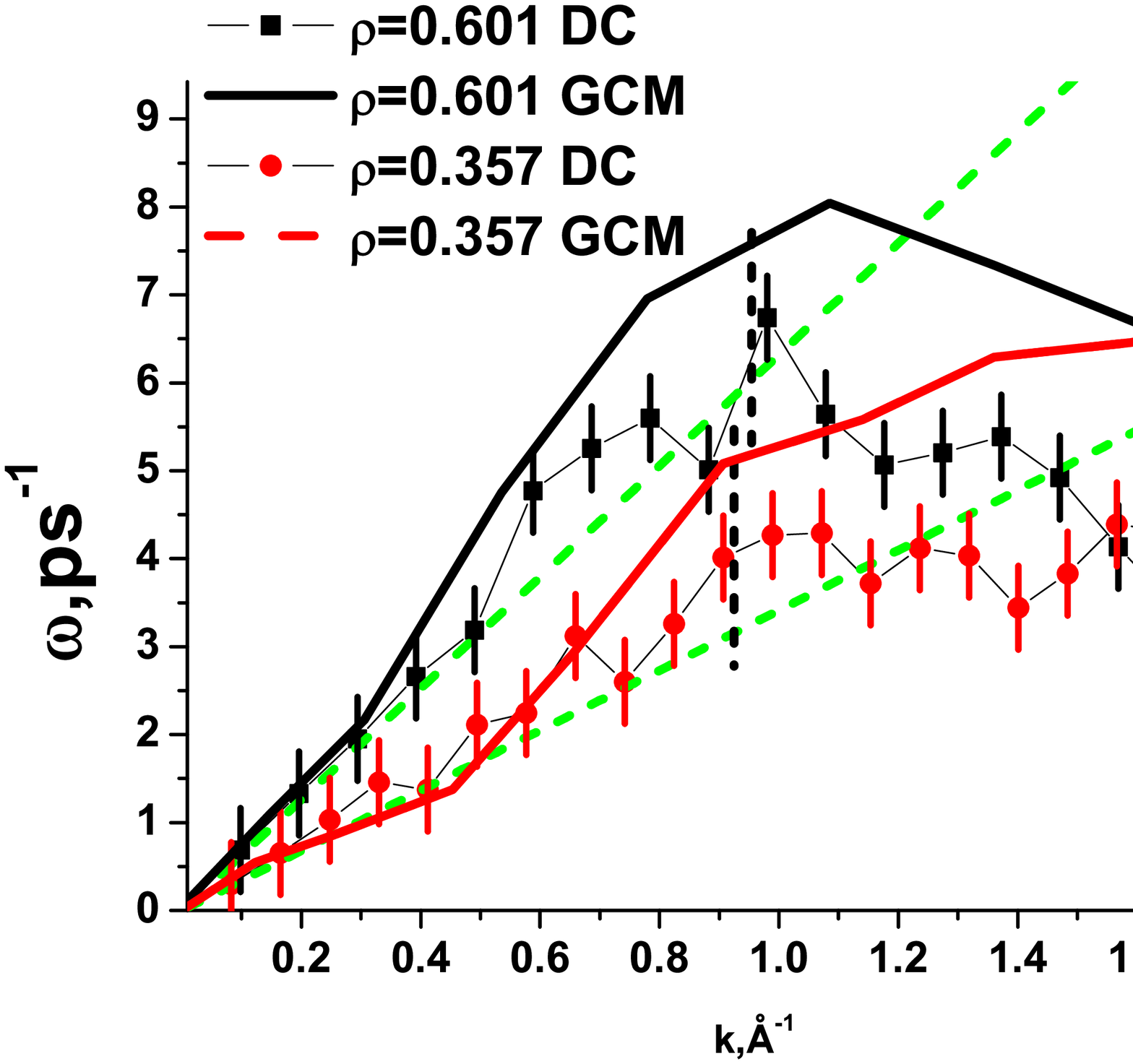}
\includegraphics[width=5cm, height=5cm]{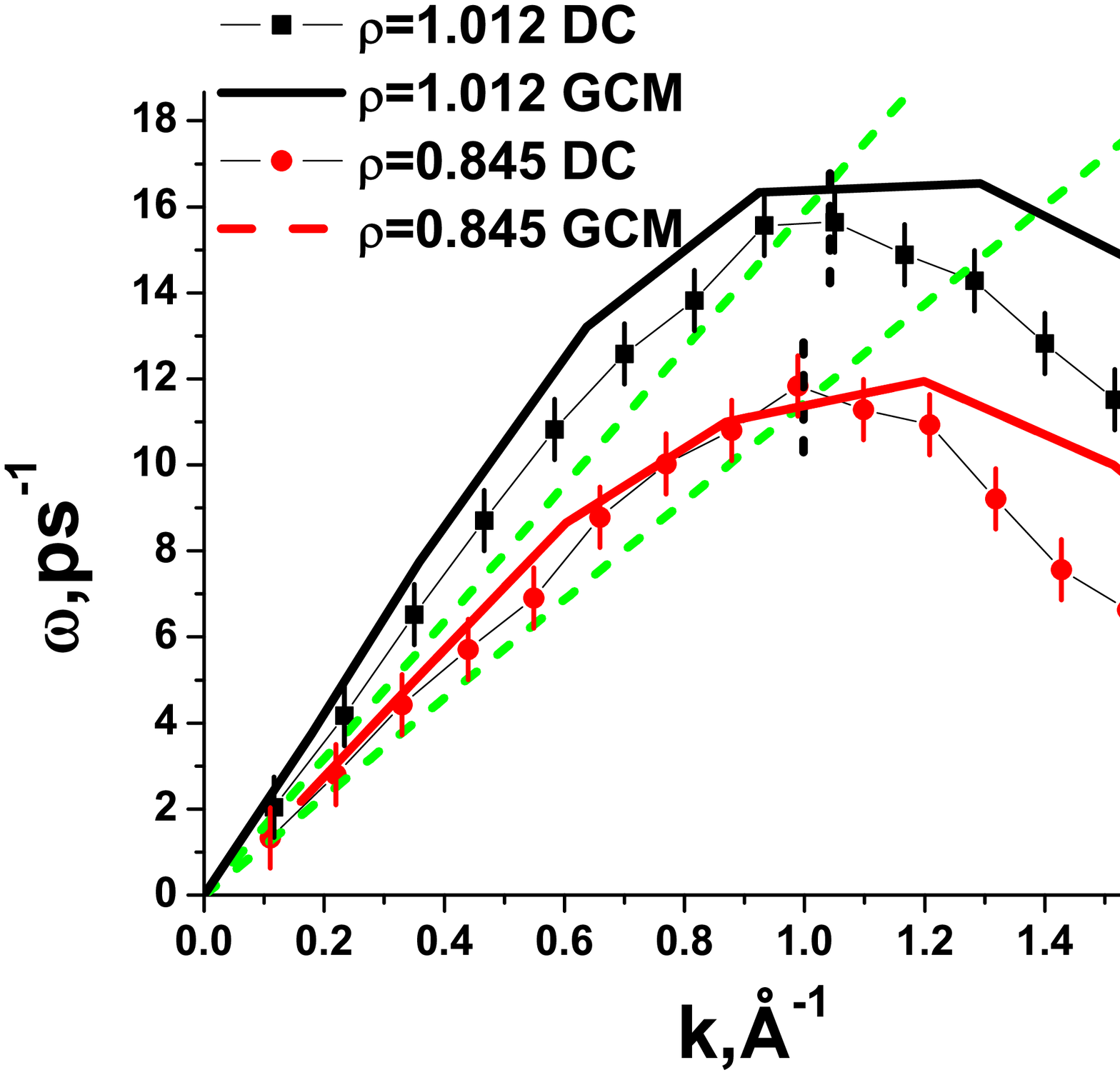}
\caption{Dispersion curves of LJ fluid (a) above and (b) below the
Frenkel line at $T=1.71$. Results from direct MD calculations and
GCM approximation from \cite{bryk-lj} are shown. Short dashed
lines crossing the data curves show the location of
pseudo-Brillouin zone. The straight lines show the Debye
dispersion law $\omega = c_s k$ where the adiabatic speed of sound
is taken from the NIST database \cite{nist}.} \label{long-t171}
\end{figure}

Our observation regarding propagating modes is related to the
important point of their detection. Theoretically, solid-like
modes were predicted by simply observing that the system behaves
like a solid above the frequency $\frac{1}{\tau}$ \cite{frenkel}.
Therefore, we have detected propagating waves by direct analysis
of the current correlation function, the procedure similar to that
employed in the experiments. A notably different result can be
found if a scheme extrapolating the hydrodynamic into the
solid-like regime is employed instead. To compare the direct
results with one such scheme, the generalized collective modes
(GCM) approach \cite{bryk-pre}, we plot the dispersion curves at
the same temperature and density conditions as in the GCM study
(see Fig. 7 in Ref. \cite{bryk-lj}) in Fig. \ref{long-t171}. We
observe that the GCM extrapolation gives reasonable results for
small $k$ only but not for larger $k$: the GCM systematically
overestimates the excitations frequency in both rigid and
non-rigid regimes. As a result, the unphysical PSD in the non-rigid fluid
is observed in the GCM calculation. A known disadvantage of the generalized hydrodynamics
is related to the phenomenological and empirical nature of the method \cite{pilgrim2,mon-na},
the general problem that involves more specific issues in the case of the GCM scheme. The choice
of variables in the GCM method is largely ad-hoc. Next, the GCM method is internally
inconsistent and violates the sum rules because the sum rules can only be exactly fulfilled if an
infinite number of exponentials is used (see also \cite{summ-rule}). Finally, the GCM
extrapolation incorrectly describes the important region of large $k$ including the
crossover to the single-particle regime (see, for example, eq. 21 in \cite{ruocco}). This is apparent in Figure \ref{long-t171}
showing failure of the GCM method at large $k$.

In summary, we have addressed the propagation of collective
excitations in the supercritical state, and directly detected
their strong crossover. We have found that the supercritical
system sustains propagating solid-like transverse mode below the
Frenkel line but becomes devoid of transverse modes above the line
where it supports longitudinal mode only. Our results
quantitatively explain the phenomenon of PSD originating from
transverse modes in the supercritical state below the Frenkel
line. We have directly ascertained that the line demarcates the
supercritical phase diagram into two regions where PSD does and
does not operate.

\begin{acknowledgments}
Yu. F. thanks the Russian Scientific Center at Kurchatov Institute
and Joint Supercomputing Center of Russian Academy of Science for
computational facilities. Yu. F., E. Ts., V. R. and V. B. are
grateful to the Russian Science Foundation (Grant No 14-22-00093).
\end{acknowledgments}

\end{document}